\documentclass[english, aps,prd,twocolumn,twoside,superscriptaddress,floatfix, nofootinbib]{revtex4-1}

\usepackage[utf8]{inputenc}
\usepackage{amsmath}
\usepackage{geometry}
\usepackage[dvipsnames]{xcolor}
\usepackage{graphicx}
\usepackage{float}
\usepackage{lipsum, babel}
\usepackage{fontawesome5}

\usepackage{tensor}
\usepackage{graphicx}
\usepackage{color}
\usepackage{hyperref}
\usepackage{ifthen,amsmath,amssymb, bm}
\usepackage{orcidlink}

\newcommand{\beq} {\begin{equation}}
\newcommand{\eeq} {\end{equation}}
\newcommand{\bal} {\begin{aligned}}
\newcommand{\eal} {\end{aligned}}

\begin{document}

\title{Evolution of structure growth during dark energy domination: Insights from the cross-correlation of DESI galaxies with CMB lensing and galaxy magnification}

\author{Noah Sailer}
\email{nsailer@berkeley.edu}
\affiliation{Berkeley Center for Cosmological Physics, Department of Physics,
University of California, Berkeley, CA 94720, USA}
\affiliation{Lawrence Berkeley National Laboratory, One Cyclotron Road, Berkeley, CA 94720, USA}

\author{Joseph DeRose}
\affiliation{Physics Department, Brookhaven National Laboratory, Upton NY 11973, USA}
\affiliation{Lawrence Berkeley National Laboratory, One Cyclotron Road, Berkeley, CA 94720, USA}

\author{Simone Ferraro}
\affiliation{Lawrence Berkeley National Laboratory, One Cyclotron Road, Berkeley, CA 94720, USA}
\affiliation{Berkeley Center for Cosmological Physics, Department of Physics,
University of California, Berkeley, CA 94720, USA}

\author{Shi-Fan Chen}
\affiliation{School of Natural Sciences, Institute for Advanced Study, 1 Einstein Drive, Princeton, NJ 08540, USA}

\author{Rongpu Zhou}
\affiliation{Lawrence Berkeley National Laboratory, One Cyclotron Road, Berkeley, CA 94720, USA}

\author{Martin White}
\affiliation{Berkeley Center for Cosmological Physics, Department of Physics,
University of California, Berkeley, CA 94720, USA}
\affiliation{Lawrence Berkeley National Laboratory, One Cyclotron Road, Berkeley, CA 94720, USA}

\author{Joshua Kim}
\affiliation{Department of Physics and Astronomy, University of Pennsylvania, 209 South 33rd Street, Philadelphia, PA 19104, USA}

\author{Mathew Madhavacheril}
\affiliation{Department of Physics and Astronomy, University of Pennsylvania, 209 South 33rd Street, Philadelphia, PA 19104, USA}

\begin{abstract}
We use a Hybrid Effective Field Theory (HEFT) model to constrain the evolution of low-redshift $(z\lesssim0.4)$ matter fluctuations by cross-correlating DESI Bright Galaxy Survey (BGS) legacy imaging with the latest CMB lensing maps from \textit{Planck} and ACT. Our tomographic BGS analysis finds that the evolution and amplitude of matter fluctuations align with CMB-conditioned $\Lambda$CDM predictions. When including DESI Baryon Acoustic Oscillation (BAO) measurements we obtain $\sigma_8 = 0.876^{+0.051}_{-0.067}$ from BGS alone. Jointly analyzing BGS and Luminous Red Galaxy (LRG) cross-correlations with the same CMB lensing maps yields $\sigma_8 = 0.791\pm0.021$. As a complementary approach we isolate the galaxy magnification signal from the cross-correlation of non-overlapping BGS and LRG photometric redshift bins, ruling out the null-magnification hypothesis at $11\sigma$. For the first time, we constrain quasi-linear structure growth from the (finite-difference calibrated) galaxy magnification signal and find $\sigma_8=0.720\pm0.047\,\,({\rm stat.})\pm0.050\,\,({\rm sys.})$ assuming a linear bias model and including BAO data.
\end{abstract}

\maketitle

\section{Introduction}
\begin{figure}[!h]
    \centering
    \includegraphics[width=\linewidth]{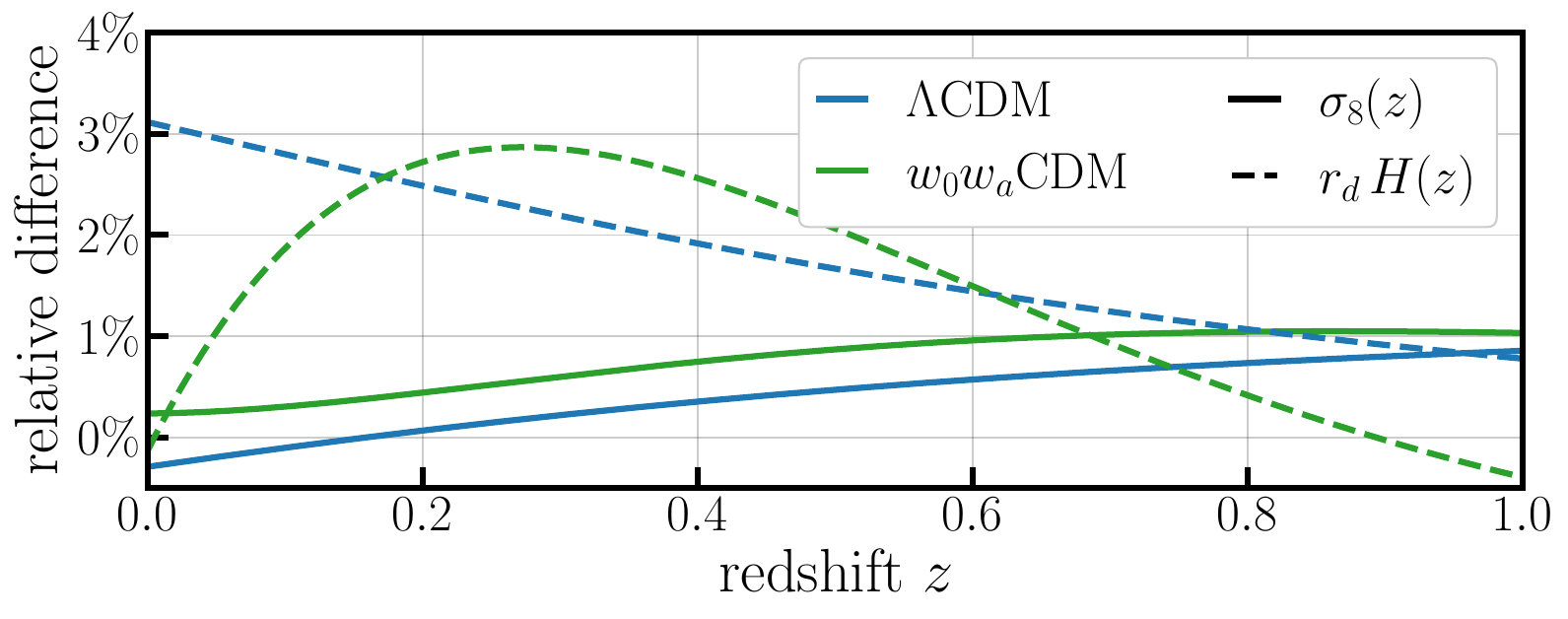}
    \caption{Relative difference in the growth (solid) and expansion (dashed) history between best-fit DESI $\Lambda$CDM (blue) or $w_0w_a$CDM (green) \cite{DESIY3BAO} and best-fit \textit{Planck}+ACT $\Lambda$CDM \cite{Planck:2018vyg,Louis:2025tst} cosmologies.
    Curves are computed with \texttt{CLASS} \cite{CLASS}.
    }
    \label{fig:rel_dif_hz_sig8}
\end{figure}

The linear growth of matter perturbations is a precise prediction of the standard cosmological model. 
While variations in the dark energy (DE) equation of state are imprinted on the low-redshift growth history via Hubble drag, in practice this effect is small relative to $\delta H/H$ 
for the cosmologies favored by the DESI Y3 BAO data \cite{DESIY3BAO} (see Fig.~\ref{fig:rel_dif_hz_sig8}).
Thus the low-redshift evolution of $\sigma_8(z)/\sigma_8$ offers a complementary view of DE which is instead primarily sensitive to DE interactions \cite[e.g.][]{Skordis:2015yra,Kumar:2017bpv,Asghari:2019qld,DiValentino:2019jae,Lucca:2021dxo,Archidiacono:2022iuu,Poulin:2022sgp,2024PhRvD.110b3536L} or modified gravity alternatives \cite{Jain10,Joyce15,Joyce16,EUCLID18,Slosar19c}, while a comparison of the present-day $\sigma_8$ with the primordial amplitude $A_s$ probes e.g. the neutrino mass sum \cite{PhysRevLett.45.1980,Hu:1997mj,Lesgourgues:2012uu,Green:2024xbb,Loverde:2024nfi} or deviations in the expansion history at higher redshifts \cite{Chen20b,Sailer:2021yzm}. 

At low redshifts where DE is relevant the universe exhibits large nonlinearities, limiting the statistical precision achievable from perturbative scales. Higher precision can be attained by incorporating smaller-scale measurements such as galaxy shear \cite{Heymans:2020gsg,KiDS:2021opn,DES:2021wwk,DES:2021bvc,DES:2021vln,Miyatake:2023njf,Sugiyama:2023fzm,Dalal:2023olq,Li:2023tui} but this comes at the cost of modeling increasingly complex and non-linear processes. 
Accurately modeling these processes is essential for robust inference. Recognition of systematic effects -- such as baryonic feedback \cite{AtacamaCosmologyTelescope:2020wtv,DES:2024iny,Hadzhiyska:2024qsl} and intrinsic alignments \cite{Schmitz:2018rfw,Blazek:2017wbz,Vlah:2019byq,Chen:2023yyb,Chen2024} -- has led to greater reported uncertainties from these probes, bringing them to a level comparable to those obtainable from perturbative scales.

In this work we constrain $\sigma_8(z)$ using two approaches: (1) tomographic cross-correlations of DESI imaging data with CMB lensing and (2) the correlation of galaxy positions with galaxy magnification. While the former is more robust to non-linear systematic effects, the latter serves as a new probe of the low-redshift clustering on large scales that is complementary to CMB lensing and galaxy shear measurements\footnote{See \cite{Xu:2024cix} for an alternative approach correlating source fluxes (rather than positions) with the lens sample on highly non-linear scales ($k\lesssim50\,\,h\,{\rm Mpc}^{-1})$.}. We describe the data and models used in each of these analyses in \S\ref{sec:data} and \S\ref{sec:likelihood} respectively. Our results are presented in \S\ref{sec:results} and we conclude in \S\ref{sec:discussion_and_conclusions}. 

\section{Data}
\label{sec:data}

We use both the Bright Galaxy Survey (BGS) \cite{Hahn:2022dnf} and Luminous Red Galaxy (LRG) \cite{DESI:2022gle,Zhou:2023gji} samples from the DESI Legacy Imaging Surveys DR9 \cite{2019AJ....157..168D} alongside the latest CMB lensing maps from \textit{Planck} PR4 \cite{Carron:2022eyg} and ACT DR6 \cite{ACT:2023dou,ACT:2023ubw,ACT:2023kun}.
The BGS and LRG imaging data cover 18200 deg$^2$ with mean angular number densities of 950 and 500 deg$^{-2}$ respectively,
while the CMB lensing data are signal dominated per mode out to $\ell\simeq80$ (\textit{Planck}) and $\ell\simeq200$ (ACT). 
The imaging data have 16600 and 7900 deg$^2$ of overlap with the \textit{Planck} and ACT lensing footprints, yielding 25 and 50$\sigma$ detections of a non-zero CMB lensing cross-correlation for the BGS and LRG samples on the scales used in our analysis.

Ref.~\cite{Zhou:2023gji} divides the LRG sample into four photometric redshift bins whose redshift distributions are calibrated using 2.3 million DESI redshifts, while ref.~\cite{Chen2024} splits the BGS sample into two spectroscopically calibrated photo-$z$ bins following an analogous approach. The redshift distributions of the six photo-$z$ bins are plotted in purple in Fig.~\ref{fig:sigma8z_wBGS}.
Number count slopes $(s_\mu \equiv d\log_{10} N/dm)$ for each bin were computed via finite differences following ref.~\cite{Zhou:2023gji} and are listed in Table~\ref{tab:sample_properties}.
An extensive set of systematics checks were performed for the LRGs by refs.~\cite{DESI:2022gle,Zhou:2023gji,2024JCAP...12..022K,2024arXiv240704607S} which found negligible systematic error at the level of cosmological parameters, and a subset of these tests for the BGS sample were performed by \cite{Chen2024}.
Refs. \cite{ACT:2024npz,Qu:2024sfu} estimate that with (tSZ) profile-hardening \cite{Sailer:2020lal,Sailer:2022jwt} residual extragalactic foreground contamination in the ACT DR6 $\kappa$ map yield $\leq0.02\sigma$ biases when cross-correlating with the BOSS LOWZ and Hang et al. \cite{Hang:2020gwn} samples, which together cover the BGS redshift range.
We deem those tests sufficient for our purposes, and further note that the BGS sample is notably brighter (hence smaller systematic modulation is expected) and is $\simeq3\times$ less constraining than the LRGs. 

\begin{figure}[!h]
    \centering
    \includegraphics[width=\linewidth]{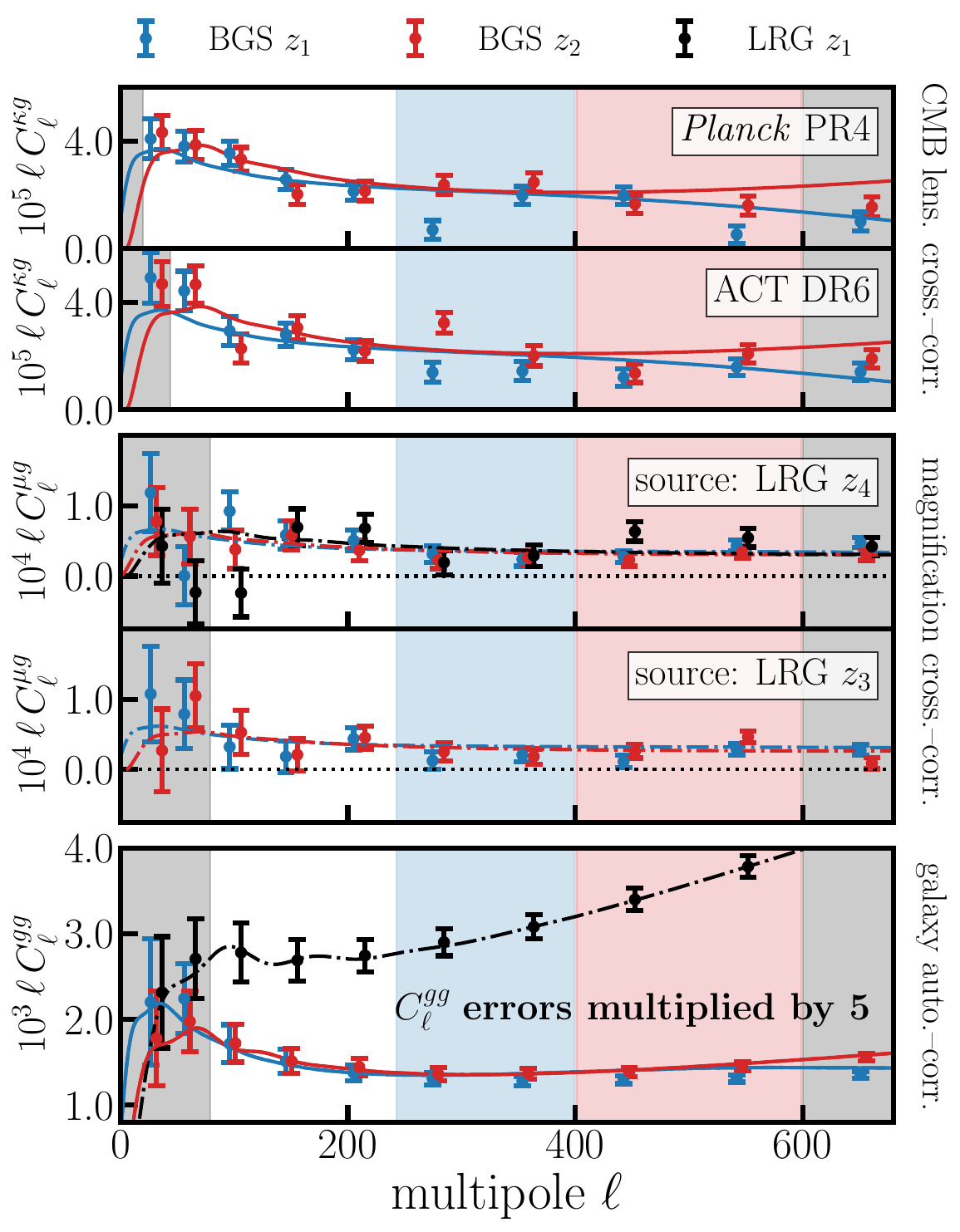}
    \caption{We show CMB lensing cross-correlations $(C^{\kappa g}_\ell)$ in the top two panels, lens-source cross-correlations $(C^{\mu g}_\ell)$ in the third and fourth panels, and galaxy auto-correlations $(C^{gg}_\ell)$ in the bottom panel. Colors indicate the galaxy sample. Black shaded regions are excluded by our scale cuts for all three samples; red regions are additionally excluded for both BGS bins; while the blue regions are additionally excluded for the lowest redshift BGS bin.
    Solid (dot dashed) lines indicate best-fit predictions from our BGS--CMB lensing cross-correlation (galaxy magnification) analysis.
    Note that the BGS best-fit $C^{gg}_\ell$ curves from both analyses lie on top of one another. We have shifted the data (and best-fit curves) by $\Delta\ell=\pm5$ and rescaled the $C^{gg}_\ell$ errors by a factor of 5 for clarity.
    }
    \label{fig:bgs_ckg_cgg}
\end{figure}

We use \texttt{NaMaster} \href{https://github.com/LSSTDESC/NaMaster}{\faGithub} \cite{Alonso:2018jzx} to measure angular power spectra and compute their associated window functions. 
We adopt the same bandpowers and analytic approximation to the covariance as in ref.~\cite{2024arXiv240704607S}. We apply a normalization correction to the raw CMB lensing cross-correlation measurements to approximately account for the nontrivial mode couplings introduced by masks and filters \cite{ACT:2023oei,Sailer2024}.
In Fig.~\ref{fig:bgs_ckg_cgg} we show the cross-correlation of the two BGS photo-$z$ bins with CMB lensing (first two panels), the cross-correlation of BGS and the lowest redshift LRG bin (serving as lenses) with the two highest redshift LRG bins (serving as sources; third and fourth panels) and the BGS auto-correlations (bottom panel). 
On small scales we adopt an $\ell_{\rm max}$ such that the approximate $k_{\rm max}=0.6\,\,h\,{\rm Mpc}^{-1}$ at the effective redshift of each photo-$z$ bin. On these scales we expect our fiducial model (\S\ref{sec:likelihood}) to be accurate to subpercent precision \cite{Kokron21}.
On large scales we choose $\ell_{\rm min}$ such that both beyond-Limber effects (including redshift-space distortions) and the residual mean field contributions to the CMB lensing estimator are negligible \cite{2024arXiv240704607S}.  

Our measurements are primarily sensitive to $S_8\equiv\sigma_8\sqrt{\Omega_m/0.3}$. To break the degeneracy between $\sigma_8$ and $\Omega_m$ and to mitigate volume effects we include the DESI Year 1 BAO data \cite{DESI:2024mwx} which effectively acts as a prior on the present day matter density $(\Omega_m=0.295\pm0.015)$.

\setlength{\tabcolsep}{4pt} % Default value: 6pt
\renewcommand{\arraystretch}{1.3} % Default value: 1
\begin{table*}[t]
    \centering
    \resizebox{\textwidth}{!}{\begin{tabular}{c|ccc|cccc|ccccccc}
         \multicolumn{4}{l}{\textbf{sample}\,\,\,\,\,\,\textbf{physical properties}} & \multicolumn{4}{c}{\textbf{scale cuts}}&\multicolumn{7}{c}{\textbf{priors on nuisance parameters}}\\
         & $z_{\rm eff}$ & $\bar{s}_\mu$ & $10^7\,{\rm SN}^{\rm 2D}_{\rm pois}$ & $\ell^{\rm auto}_{\rm min}$ & $\ell^{\rm \textit{Planck}}_{\rm min}$ & $\ell^{\rm ACT}_{\rm min}$ & $\ell_{\rm max}$ & $b_1^L(z_{\rm eff})$ & $b_2^L(z_{\rm eff})$ & $b_s^L(z_{\rm eff})$ & $\alpha_a(z_{\rm eff})$ & $\alpha_x(z_{\rm eff})$ & ${\rm SN}^{\rm 2D}$ & $s_\mu$\\
         \hline
         BGS $z_1$ & 0.211 & 0.81 & 4.63 & 79 & 20 & 44 & 243 & $\mathcal{U}(-1,2)$ & $\mathcal{U}(-5,5)$ & $\mathcal{N}(0,1)$ & $\mathcal{N}(0,50)$ & $\mathcal{N}\Big(\frac{\alpha_a}{2b_1^E}\,,\,2\Big)$ & $\mathcal{N}_r({\rm SN}^{\rm 2D}_{\rm pois},0.3)$ & $\mathcal{N}(\bar{s}_\mu,0.1)$\\
         BGS $z_2$ & 0.352 & 0.80 & 9.18 & 79 & 20 & 44 & 401 & \multicolumn{7}{c}{\dotfill\dotfill\quad\quad same as above\quad\quad\dotfill\dotfill}\\
    \end{tabular}}
    \caption{
    A summary of the BGS physical properties (effective redshift, magnification bias, Poisson shot noise), scale cuts and priors on nuisance parameters for our tomographic CMB lensing analysis. 
    $\mathcal{N}(\mu,\sigma)$ and $\mathcal{U}({\rm min},{\rm max})$ indicate Gaussian and uniform distributions respectively, while $\mathcal{N}_r(\mu,\sigma/\mu) \equiv \mathcal{N}(\mu,\sigma)$. The linear Eulerian bias is given by $b_1^E = 1+b_1^L$.
    }
    \label{tab:sample_properties}
\end{table*}

\section{Likelihoods}
\label{sec:likelihood}

Our analysis of the CMB lensing cross-correlation with the BGS sample (\S\ref{sec:cmb_lens_tomography}) closely mirrors the approach taken by refs.~\cite{2024JCAP...12..022K,2024arXiv240704607S} for the DESI LRGs, for which the code (\verb|MaPar| \href{https://github.com/NoahSailer/MaPar/tree/main}{\faGithub}) is publicly available. The likelihood used for our magnification-derived growth constraints (\S\ref{sec:mag_constraints}) follows a nearly identical structure but with the lens-source cross-correlation replacing the CMB lensing cross-correlation. 
Below we briefly summarize the model and priors adopted for the BGS and LRG samples.
We then show that our analysis choices yield acceptably small ``volume effects" and  faithfully recover the input cosmology to a set of BGS-like simulations.

\subsection{Model, priors and sampling}
\label{sec:model_prior_sampling}

We model the galaxy density contrast using a Hybrid Effective Field Theory (HEFT) \cite{Modi:2019qbt,Kokron21,Hadzhiyska21,Zennaro:2021pbe,Pellejero-Ibanez:2022efv,Nicola:2023hsd}
implemented with the \texttt{Aemulus} $\nu$ simulations \href{https://github.com/AemulusProject/aemulus_nu_public}{\faGithub} \cite{DeRose:2023dmk}.
This implementation includes linear ($b_1^L$), quadratic ($b^L_2$ and $b^L_s$) and leading-order derivative ($b^L_{\nabla^2}$) contributions to the galaxy bias expansion which enables a rigorous marginalization over the small-scale astrophysics relevant for galaxy formation.
We use the Limber approximation \cite{1953ApJ...117..134L,1992ApJ...388..272K,LoVerde:2008re} to compute the galaxy auto-correlation $(C^{gg}_\ell)$ and cross-correlation $(C^{\kappa g}_\ell)$ with CMB lensing. We adopt an effective redshift $(z_{\rm eff})$ approximation when evaluating the three dimensional galaxy-galaxy $(P_{gg})$ and galaxy-matter $(P_{gm})$ power spectra within the Limber integrals.
Due to the BGS sample's narrow $dN/dz$'s (see Fig.~\ref{fig:sigma8z_wBGS}) the effective redshift approximation has subpercent accuracy for physically reasonable galaxy bias evolutions. 
The effective redshift approximation dramatically simplifies the modeling, since bias parameters need only to be specified at each $z_{\rm eff}$ rather than their evolution within each photo-$z$ bin.
For additional details about the model and its implementation, including the treatment of pixelization and magnification, we refer the reader to \S4 of ref.~\cite{2024arXiv240704607S}. 

We model the cross-correlation of source and lens galaxies following \S6.6 of ref.~\cite{2024arXiv240704607S}.
For simplicity, we have chosen only to include lens-source pairs with negligible redshift overlap in our magnification-derived growth constraints.
In this case, the cross-correlation between the source sample $\mu$ and the lens $g$ is very well approximated by
\begin{equation}
\label{eq:source-lens}
    C^{\mu g}_\ell
    \simeq 
    \int\frac{d\chi}{\chi^2}
    W^\mu(\chi) W^g(\chi)
    P_{gm}\Big(k=\frac{\ell+1/2}{\chi},z_{\rm eff}\Big),
\end{equation}
where $W^\mu \propto (5s_\mu-2)$ is the magnification kernel of the source, $W^g\propto dN/dz$ is the projection kernel of the lens and $z_{\rm eff}$ is the effective redshift of the lens. 
Provided that the magnification bias $s_\mu$ of the source sample is known more precisely than the amplitude of the lens-source cross-correlation one can heuristically measure $S_8$ from the ratio $C^{\mu g}_\ell/\sqrt{C^{gg}_\ell}$.
We use Eq.~\eqref{eq:source-lens} as our lens-source model in \S\ref{sec:mag_constraints} to derive growth constraints from the galaxy data alone.

We adopt a Gaussian likelihood and analytically marginalize over nuisance parameters which enter linearly into the model (see Appendix D of \cite{2024arXiv240704607S}) to decrease the convergence time of our chains. Posteriors are sampled with \verb|cobaya|'s \cite{Torrado:2020dgo,2019ascl.soft10019T} Markov Chain Monte Carlo Metropolis sampler \cite{Lewis:2002ah,Lewis:2013hha}. 
Plots of marginal distributions are obtained with the \verb|GetDist| \cite{Lewis:2019xzd} package.
We estimate best-fit values using \verb|cobaya|'s default minimizer.

As in \cite{2024JCAP...12..022K,2024arXiv240704607S} we fix $n_s=0.9649$, $\Omega_b h^2=0.02236$, $\Omega_m h^3=0.09633$ and $N_{\rm eff} = 3.044$ to their \textit{Planck} 2018 mean values\footnote{Our results are largely insensitive to $n_s$: increasing $n_s$ from 0.965 to 0.984 increases the inferred $S_8$ from our baseline analysis by 0.003 $(\simeq0.14\sigma)$.} \cite{Planck:2018vyg} and fix the neutrino mass sum to $M_\nu = 60$ meV. We sample $\ln(10^{10}A_s)$ and $\Omega_c h^2$ with uniform $\mathcal{U}(2,4)$ and $\mathcal{U}(0.08,0.16)$ priors respectively.
We place uniform $\mathcal{U}(-1,2)$ priors on the linear (Lagrangian) bias parameters ($b_1^L$) for both BGS photo-$z$ bins. 
Priors on higher order bias parameters, counterterms, and shot noise are set in an identical fashion to those used in the fiducial analysis of refs.~\cite{2024JCAP...12..022K,2024arXiv240704607S} and are listed in Table \ref{tab:sample_properties}. 
We marginalize over the magnification bias with the priors listed in Table~\ref{tab:sample_properties} for CMB lensing-derived growth constraints (\S\ref{sec:cmb_lens_tomography}) while for magnification-derived constraints (\S\ref{sec:mag_constraints}) we hold magnification bias fixed to the central values measured by refs.~\cite{Zhou:2023gji,Chen2024}.

\subsection{Likelihood tests}
\label{sec:likelihood_tests}

\textbf{CMB lensing tomography:} To estimate the expected size of volume (projection) effects in our CMB lensing analysis we analyze a noiseless data vector generated from our fiducial model with cosmological parameters fixed to those used for the \texttt{Buzzard} simulations (in particular, $S_8=0.801$ and $\sigma_8=0.820$). We assume a linear bias of $b_1^L(z_{\rm eff}) = 0.0,\,0.2$ \cite{Chen2024} for the two redshift bins, use the measured magnification bias and shot noise listed in Table~\ref{tab:sample_properties} (under ``physical properties") and set all higher order biases to $0$ in our fiducial model prediction. 
Throughout we use the same covariance here as in our fiducial BGS analysis.

We list marginal $S_8$ posteriors obtained from the synthetic cross- and auto-correlation alone (i.e. without BAO information) in the first column of Table~\ref{tab:volume_and_buzzard}. 
We find that when fitting to both BGS bins the mean $S_8$ is $\simeq 0.8\sigma$ lower than the input value.
The offsets are smaller when fitting to the individual samples: $\simeq 0.5\sigma$ and $\simeq0.2\sigma$ for the lower and higher redshift bins respectively.

One could in principle mitigate volume effects by adopting informative priors on galaxy nuisance parameters. As an extreme case we fixed magnification bias and all higher order contributions to their true values so that the only nuisance parameters varied are linear bias and shot noise, and still found the mean $S_8$ to be $\simeq0.4\sigma$ lower than the input value when fitting to both BGS bins. 
An alternative approach to mitigate these shifts is to include BAO data when quoting BGS-only $S_8$ constraints. To demonstrate this we generated a noiseless DESI Y1 BAO data vector evaluated at the \texttt{Buzzard} cosmology. When including the mock BAO data (second and third columns of Table~\ref{tab:volume_and_buzzard}) we find that mean $S_8$ and $\sigma_8$ values are within $0.25\sigma$ of their input values when analyzing the BGS bins individually and together; thus we always include BAO in \S\ref{sec:results} for both our $S_8$ and $\sigma_8$ constraints.

\setlength{\tabcolsep}{4pt} % Default value: 6pt
\renewcommand{\arraystretch}{1.3} % Default value: 1
\begin{table}[!h]
    \centering
    \resizebox{0.5\textwidth}{!}{\begin{tabular}{c||ccc}
    & $S_8$ without BAO & $S_8$ with BAO & $\sigma_8$ with BAO\\
    \hline
    \hline
    Fid. value & 0.801 & 0.801 & 0.820\\
    \hline
    \hline
    & \multicolumn{3}{c}{\textbf{noiseless model prediction}} \\
    BGS $z_1$ & $0.782^{+0.050}_{-0.092}$ $[0.806]$ & $0.800^{+0.067}_{-0.098}$ $[0.798]$ & $0.819^{+0.069}_{-0.100}$ $[0.817]$\\
    BGS $z_2$ & $0.792^{+0.049}_{-0.065}$ $[0.793]$ & $0.808^{+0.048}_{-0.071}$ $[0.826]$ & $0.827^{+0.050}_{-0.073}$ $[0.836]$\\
    BGS ($z_1$ \& $z_2$) & $0.765^{+0.041}_{-0.051}$ $[0.802]$ & $0.793^{+0.044}_{-0.054}$ $[0.798]$ & $0.810^{+0.047}_{-0.057}$ $[0.816]$\\
    \hline
    & \multicolumn{3}{c}{\textbf{\texttt{Buzzard} simulations}}\\
    BGS $z_1$ & $0.776^{+0.052}_{-0.089}$ $[0.803]$ & $0.800^{+0.060}_{-0.092}$ $[0.788]$ & $0.819^{+0.062}_{-0.095}$ $[0.805]$\\
    BGS $z_2$ & $0.781^{+0.042}_{-0.055}$ $[0.806]$ & $0.766^{+0.043}_{-0.065}$ $[0.787]$ & $0.783^{+0.046}_{-0.068}$ $[0.804]$\\
    BGS & $0.740^{+0.039}_{-0.050}$ $[0.790]$ & $0.762^{+0.040}_{-0.050}$ $[0.801]$ & $0.778^{+0.042}_{-0.053}$ $[0.819]$\\
    \end{tabular}}
    \caption{Tomographic CMB lensing constraints (formatted as: mean $\pm1\sigma$ $[$best fit$]$) from our fits to noiseless synthetic data (top three rows) and our fits to the \texttt{Buzzard} mock measurements (bottom three rows).
    In all cases we use the same covariance as in our fiducial analysis.
    The scatter in best-fits around the fiducial values for the noiseless model prediction are representative of the minimizer's precision.
    }
    \label{tab:volume_and_buzzard}
\end{table}

To test our likelihood in a more realistic setting we next analyze a mock BGS sample built on the \texttt{Buzzard v2.0} suite of $N$-body simulations. We refer the reader to \cite{DeRose19,DeRose25} and the references therein \cite{Springel:2005mi,DES:2021bwg,DeRose:2021avs,Wechsler:2021esl} for a detailed description of the sample's construction.
The galaxy auto- and cross-correlation with CMB lensing are measured on seven quarter-sky simulations and binned into bandpowers in an identical fashion to the data. 
When fitting to these mock measurements we adopt the same scale cuts as in \S\ref{sec:data}, use the mock window functions and redshift distributions, and adjust the fiducial values of fixed cosmological parameters to match those used in the \texttt{Buzzard} simulations, while the priors for sampled parameters remain identical to those listed in Table~\ref{tab:sample_properties}. 

The \texttt{Buzzard} simulations cover $\simeq4\times$ the area of the imaging data. Thus $\simeq0.5\sigma$ fluctuations about the input cosmology are expected when fitting to the \texttt{Buzzard} measurements with the covariance used in our fiducial BGS analysis. 
To maintain this expectation when including BAO we add Gaussian noise to the mock BAO data described above whose covariance is $1/4\times$(the DESI Y1 BAO covariance). Results of the \texttt{Buzzard} tests are listed in the last three rows of Table~\ref{tab:volume_and_buzzard}.
When fitting to the first (second) BGS bin only all of the posteriors are within $0.5\sigma$ ($0.7\sigma$) of the input cosmology.
When jointly analyzing both bins we find $0.9\sigma$ ($1.3\sigma$) differences with (without) the inclusion of mock BAO data.
From the fits to a noiseless model prediction we expect that $\simeq0.25\sigma$ ($\simeq0.8\sigma$) of these differences can be attributed to volume effects. 
We conclude that our mock tests are consistent within $\simeq0.7\sigma$ (or smaller) fluctuations about the input cosmology when volume effects are taken into account, which are mild $(<0.25\sigma)$ with the inclusion of BAO data.

\textbf{Galaxy magnification:} We repeated the volume effect test for our galaxy magnification analysis (\S\ref{sec:mag_constraints}) on a synthetic data vector and found $\simeq2\sigma$ shifts in $S_8$ and $\sigma_8$ for the priors used in the CMB lensing analysis (Table \ref{tab:sample_properties}), despite including synthetic BAO data and fixing magnification bias\footnote{In practice the Poisson errors of the finite difference method are much smaller ($\sim1\%$, see ref.~\cite{Zhou:2023gji}) than the precision of our magnification-derived growth constraint.}.
Given the modest ${\rm SNR}=11$ of the magnification signal we chose to instead adopt a linear bias model. For the linear bias model volume effects are at most $0.3\sigma$ when BAO data are included.

We estimate the systematic error from beyond linear biasing as follows. We fix the cosmological parameters of our joint BGS \& LRG likelihood (\S\ref{sec:cmb_lens_tomography}) to the \texttt{Buzzard} cosmology, shot noise to its Poisson estimate and number count slopes to their measured finite difference values. We then obtain best-fit values for the remaining nuisance parameters. Using these values we generate a synthetic data vector that includes realistic beyond linear biasing, which is then fit using our linear bias magnification likelihood including synthetic BAO. Doing so yields $\sigma_8=0.762\pm0.052$ $[0.776]$. Given that volume effects can account for $(\Delta\sigma_8)_{\rm vol.}=-0.015$ of this shift away from the true $\sigma_8=0.82$, we estimate a systematic shift $(\Delta\sigma_8)_{\rm sys.}\simeq-0.05$ from beyond linear biasing. 
This shift is comparable to the statistical precision and so we report both statistical and systematic errors in \S\ref{sec:mag_constraints}.

In our magnification model we neglect the redshift evolution of $s_\mu$ within the source photo-$z$ bins, which ref.~\cite{Zhou:2023gji} found to be significant for the BOSS CMASS sample. If we instead assume a linear evolution of $s_\mu(z)$ whose slope is estimated via finite differences\footnote{Specifically we set $s_\mu(z) = \bar{s}_\mu + \bar{s}'_\mu(z-\bar{z})$ for both LRG source samples, where $\bar{z}$ is the mean redshift, $\bar{s}_\mu$ is is the mean slope measured by ref.~\cite{Zhou:2023gji} and $\bar{s}_\mu'\simeq -0.4,\,1.2$ for LRG $z_3$ and $z_4$ respectively. We estimate $\bar{s}'_\mu$ by splitting each photo-$z$ bin in half and measuring $\bar{s}_\mu$ for each half.} our constraints shift by $\Delta\sigma_8=-0.002$. 
This systematic is subdominant to beyond-linear biasing and we therefore neglect it.

\section{Results}
\label{sec:results}

In \S\ref{sec:cmb_lens_tomography} and \S\ref{sec:mag_constraints} we present growth constraints from CMB lensing tomography and galaxy-magnification respectively.
To simplify the interpretation (i.e. mitigate volume effects, see \S\ref{sec:likelihood_tests}) of our results we always include DESI Y1 BAO data.
To mitigate confirmation bias in our tomographic CMB lensing analysis we added a 5\% Gaussian scatter to the mean and best-fit cosmological parameter values when producing preliminary versions of Fig.~\ref{fig:sigma8z_wBGS} until all scale cuts and priors had been finalized.

\subsection{CMB lensing tomography}
\label{sec:cmb_lens_tomography}

We summarize our growth constraints from the cross-correlation of \textit{Planck}$+$ACT CMB lensing with the DESI BGS and LRG samples in Table \ref{tab:results}. Constraints from each individual BGS photo-$z$ bin are listed in the first two rows. We jointly analyze both photo-$z$ bins to obtain the third row. For comparison we quote the joint LRG constraints \cite{2024JCAP...12..022K,2024arXiv240704607S} in the fourth row, which have been updated to include DESI Y1 BAO. In the final row we list the joint BGS \& LRG constraints.

\setlength{\tabcolsep}{4pt} % Default value: 6pt
\renewcommand{\arraystretch}{1.3} % Default value: 1
\begin{table}[!h]
    \centering
    \begin{tabular}{c||ccc}
    & $S_8$ & $\sigma_8$\\
    \hline
    \hline
    BGS $z_1$   & $0.974^{+0.072}_{-0.110}$ $[0.940]$ & $0.981^{+0.074}_{-0.110}$ $[0.946]$\\
    BGS $z_2$   & $0.785^{+0.038}_{-0.061}$ $[0.768]$ & $0.792^{+0.040}_{-0.063}$ $[0.776]$\\
    BGS         & $0.870^{+0.050}_{-0.064}$ $[0.892]$ & $0.876^{+0.051}_{-0.067}$ $[0.899]$\\
    \hline
    LRG \cite{2024JCAP...12..022K,2024arXiv240704607S} & $0.775^{+0.019}_{-0.022}$ $[0.779]$ & $0.781\pm0.022$ $[0.788]$ \\
    \hline
    BGS \& LRG & $0.788\pm0.020$ $[0.788]$ & $0.791\pm0.021$ $[0.796]$\\
    \end{tabular}
    \caption{Constraints (formatted as: mean $\pm1\sigma$ $[$best fit$]$) from cross-correlating \textit{Planck}$+$ACT CMB lensing with several combinations of the DESI BGS and LRG photo-$z$ bins. For all combinations we include DESI Y1 BAO data.}
    \label{tab:results}
\end{table}

For BGS-only (third row, Table \ref{tab:results}) we find a best-fit $\chi^2_{\kappa g,gg} = 30.2$ and $\chi^2_{\rm BAO} = 12.8$. The corresponding best-fit prediction for each BGS auto- and CMB lensing cross-correlation are plotted in Fig.~\ref{fig:bgs_ckg_cgg} (solid lines). 
With the addition of LRG data (last row, Table \ref{tab:results}) we find a best-fit $\chi^2_{\kappa g,gg} = 89.1$ and $\chi^2_{\rm BAO} = 12.8$. 
Our likelihood consists of 30 (126) data points and is fit with 16 (44) free parameters when analyzing the BGS (BGS \& LRG) data, while the DESI Y1 BAO likelihood has 12 data points. 
We emphasize that some of our free parameters (e.g. $s_\mu$, $b_s$, $\alpha_x$) are prior dominated, and as such it is nontrivial to estimate the effective number of degrees of freedom.
Instead we quote Bayesian PTEs \cite{reason:GelCarSteRub95} following Appendix B of ref.~\cite{2024arXiv240704607S}. 
For BGS (BGS \& LRG) we estimate ${\rm PTE} = 14\%\,\,($77\%$)$.

\subsection{Growth from galaxy magnification}
\label{sec:mag_constraints}

For our magnification-derived constraints we jointly analyze three galaxy auto-spectra (fifth panel of Fig.~\ref{fig:bgs_ckg_cgg}) alongside five lens-source cross-correlation measurements (third and fourth panels). The lens-source pairs have been chosen to have no redshift overlap, thus their corresponding cross-correlations vanish in the absence of magnification. From our five lens-source pairs we obtain ${\rm SNR}=11$ for the scale cuts discussed in \S\ref{sec:data} and shown in Fig.~\ref{fig:bgs_ckg_cgg}. When combined with DESI Y1 BAO these measurements yield
\begin{equation*}
\begin{aligned}
\label{eq:magnification_constraint}
    S_8 &=
    0.715\pm0.046\,\,({\rm stat.})\pm0.050\,\,({\rm sys.})\\
    \sigma_8 &=
    0.720\pm0.047\,\,({\rm stat.})\pm0.050\,\,({\rm sys.})\\
\end{aligned}
\end{equation*}
assuming a linear bias model and fixed number count slope $s_\mu$. The systematic (sys.) error quoted above comes from beyond-linear biasing and is estimated following \S\ref{sec:likelihood_tests}.
Our best-fit predictions are shown in the bottom three panels of Fig.~\ref{fig:bgs_ckg_cgg} (dot dashed). We find a best-fit $\chi^2_{\mu g,gg} = 33.8$ and $\chi^2_{\rm BAO} = 12.7$ for $38+12$ data points. We estimate the Bayesian PTE to be $40\%$. 

We only analyze far-separated lens-source pairs so that the only ``cosmological'' correlation is expected to arise from magnification. However, imaging systematics can potentially correlate bins that are disjoint in redshift; this will need to be studied and characterized for future analyses. For the present work, the modest constraining power together with the tests performed by refs.~\cite{2024JCAP...12..022K,2024arXiv240704607S,Chen2024}, which found no evidence of residual imaging systematics, lends confidence in the robustness of these results.

\begin{figure}[!h]
    \centering
    \includegraphics[width=\linewidth]{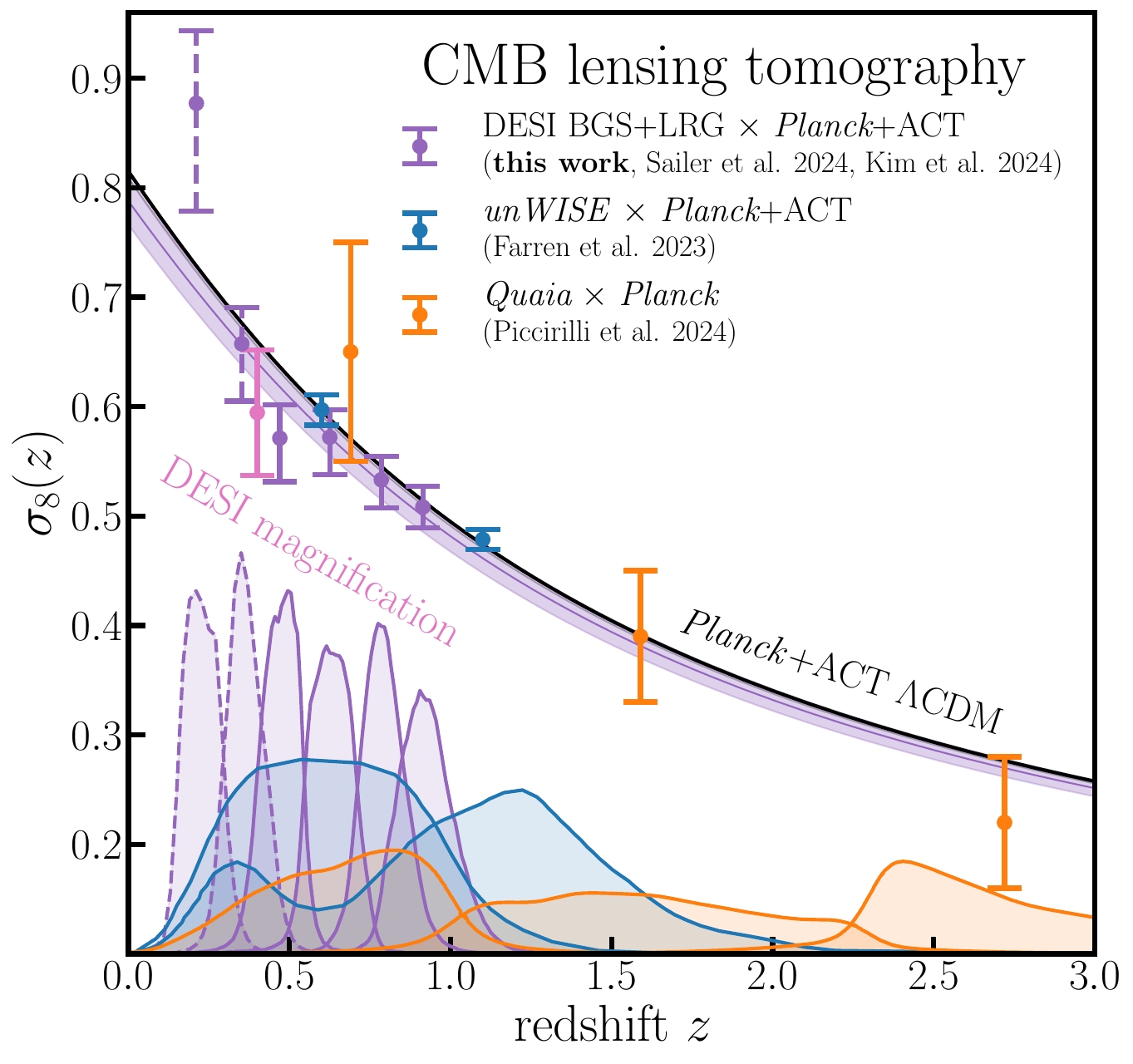}
    \caption{Structure growth from the joint analysis of BAO and CMB lensing cross-correlations with DESI BGS (dashed purple) and LRG  \cite{2024JCAP...12..022K,2024arXiv240704607S} (solid purple) imaging data, \textit{unWISE} galaxy catalogs \cite{ACT:2023oei} (blue) and \textit{Quaia} quasars \cite{Piccirilli:2024xgo} (orange).
    The shaded purple band represents the $\pm1\sigma$ credible interval for our combined BGS \& LRG fit (bottom row of Table~\ref{tab:results}).
    We plot our magnification constraint in pink and the \textit{Planck}+ACT $\Lambda$CDM evolution of $\sigma_8(z)$ in black \cite{Tristram:2023haj,Rosenberg:2022sdy,Louis:2025tst}.
    The shaded regions on the bottom of the plot are proportional to the redshift distribution of each galaxy sample.
    }
    \label{fig:sigma8z_wBGS}
\end{figure}

\section{Discussion and Conclusions}
\label{sec:discussion_and_conclusions}

We constrain the low redshift evolution of linear structure growth from the cross-correlation of DESI BGS imaging data with the latest CMB lensing maps from \textit{Planck} and ACT.
With the inclusion of BAO we find that $\sigma_8$ is consistent with CMB-conditioned $\Lambda$CDM predictions to within $\simeq 1.1\sigma$, while $S_8$ is slightly more discrepant $(\simeq1.7\sigma)$ due to the CMB preferring a higher $\Omega_m$ \cite{Planck:2018vyg,Louis:2025tst,Hang:2020gwn,Qu:2024sfu,DESIY3BAO}.
This is true both when analyzing the BGS data on their own or when jointly analyzed with the tomographic LRG measurements from refs.~\cite{2024JCAP...12..022K,2024arXiv240704607S}.
Our tomographic analysis suggest that the rate of growth at redshifts $z\lesssim1$ is also consistent with a CMB-conditioned $\Lambda$CDM cosmology. Recent CMB lensing cross-correlations with higher redshift tracers such as \textit{unWISE} galaxies \cite{ACT:2023oei} or \textit{Quaia} quasars \cite{Piccirilli:2024xgo} suggest that this trend continues to higher redshifts (see Fig.~\ref{fig:sigma8z_wBGS}).

Since low-$z$ linear growth is relatively insensitive to time-dependent DE (Fig.~\ref{fig:rel_dif_hz_sig8}) our findings do not contradict the preference for evolving DE by the DESI Y3 BAO data \cite{DESIY3BAO}. 
Our tomographic measurements provide a complementary approach to BAO observations for constraining DE properties beyond its equation of state, such as DE interactions or modified gravity alternatives to DE.

We note that our results are in tension with those recently obtained from the joint analysis of the 3D clustering of BOSS galaxies and their cross-correlation with CMB lensing \cite{Chen:2024vuf}. 
We suspect that this tension is driven by unresolved systematics in the BOSS data and we leave an in-depth investigation to future work.
In the near future, we will present a joint analysis of our measurements with the 3D clustering of DESI galaxies \cite{MM25} which will be more directly comparable to ref.~\cite{Chen:2024vuf}.

We analyze galaxy magnification as a new probe of structure growth. 
Given the modest precision of our magnification measurement, we adopt a linear bias model which yields a $\simeq 9\%$ constraint\footnote{We add the statistical and systematic errors in quadrature.} of $\sigma_8$ that is consistent with CMB-conditioned $\Lambda$CDM predictions to within $\simeq 1\sigma$. 
The precision obtainable from galaxy magnification is limited by (the cosmic variance associated with) the source sample auto-correlation, which is $\simeq10\times$ larger than the lens-source cross-correlation.
The SNR of the magnification signal may be improved by tuning the photometric selection to maximize the source number count slope while simultaneously decreasing the source linear bias.
Finally, a joint analysis of magnification and CMB lensing cross-correlations could be used to calibrate magnification bias independently of finite difference estimates, which we leave to future work. 

\section*{Acknowledgments}
We thank William Coulton, Gerrit Farren, Fiona McCarthy, Blake Sherwin and David Spergel for useful discussions during the preparation of this manuscript. 
NS and SF are supported by Lawrence Berkeley National Laboratory and the Director, Office of Science, Office of High Energy Physics of the U.S. Department of Energy under Contract No.\ DE-AC02-05CH11231. SC acknowledges the support of the National Science Foundation at the
Institute for Advanced Study through NSF/PHY 2207583.
JK and MM acknowledge support from NSF grants AST-2307727 and  AST-2153201. 
MM additionally acknowledges NASA grant 21-ATP21-0145. 

\bibliographystyle{JHEP.bst}
\bibliography{main}

\end{document}